\documentclass[a4paper]{jpconf}
\usepackage{amsmath,amssymb,amsfonts}
\usepackage{graphicx}

\def\ben{\begin{equation}}
\def\een{\end{equation}}
\def\bea{\begin{eqnarray}}
\def\eea{\end{eqnarray}}
\def\nn{\nonumber}

\def\im{{\rm i}}

\begin{document}
\title{2-point functions in quantum cosmology\footnote{Based on a talk given at Loops '11, Madrid, on 26 May 2011 \cite{talkpdf}.}}

\author{Steffen Gielen}

\address{Max Planck Institute for Gravitational Physics (Albert Einstein Institute), Am M\"uhlenberg 1, 14476 Golm, Germany}

\ead{gielen@aei.mpg.de}

\begin{abstract}
We discuss the path-integral formulation of quantum cosmology with a massless scalar field as a sum-over-histories, with particular reference to loop quantum cosmology. Exploiting the analogy with the relativistic particle, we give a complete overview of the possible two-point functions, deriving vertex expansions and composition laws they satisfy. We clarify the tie between definitions using a group averaging procedure and those in a deparametrised framework. We draw some conclusions about the physics of a single quantum universe and multiverse field theories where the role of these sectors and the inner product are reinterpreted. \end{abstract}

\section{Introduction and Motivation}

Quantum cosmology is the quantisation of dimensionally reduced models where one makes symmetry assumptions about the geometry already at the classical level. Both in the traditional approaches and in loop quantum cosmology, the motivation for studying such models has been twofold: One could hope to make predictions for cosmology, where the FRW spacetimes give a good approximation to our own Universe, or one could study mathematical and conceptual issues of quantum gravity in a simplified model where explicit calculations are possible. We focus on the second point and on the definition of the {\it physical inner product}, which for a constrained system plays the role of a transition amplitude, as a sum over histories. Using the analogy with the relativistic particle, we study different possible choices and focus on the {\it composition laws} they satisfy, in situations where a massless scalar field represents an internal time variable.

We assume a homogeneous, isotropic $k = 0$ FRW universe with a massless scalar field, with Hamiltonian constraint (here $\Theta$ can be a difference operator or a differential operator)
\ben
\hat{\mathfrak{C}}\Psi(\nu,\phi)\equiv\left(-\Theta-\partial_\phi^2\right)\Psi(\nu,\phi)=0\,,
\label{const}
\een
where $\nu$ parametrises the gravitational field and $\Theta$ is an operator acting on the gravitational part of the kinematical Hilbert space $\mathcal{H} = \mathcal{H}_{{\rm kin}}^{{\rm g}}\otimes \mathcal{H}_{{\rm kin}}^{\phi}$; $\phi$ appears as a relational time variable.

In loop quantum cosmology (LQC), one has a difference operator $\Theta$ of the form
\ben
(\Theta\Psi)(\nu,\phi)\equiv A(\nu)\Psi(\nu+\nu_0,\phi)+B(\nu)\Psi(\nu,\phi)+C(\nu)\Psi(\nu-\nu_0,\phi)\,,
\een
where the functions $A, B, C$ depend on the quantisation scheme; there is a superselection of subspaces $\{|\nu' + n\nu_0\rangle |n \in \mathbb{Z}\}$ and one may restrict wave functions to $\nu_0 \mathbb{Z}$. The solutions to (\ref{const}) can be split into positive and negative frequency parts satisfying $(\hat{p}_{\phi}\mp\sqrt{\Theta})\Psi_\pm(\nu,\phi)=0$.

In \cite{ach}, a spin foam-type expansion was done for the inner product in loop quantum cosmology. Starting from a slight modification of the group-averaging expression used in timeless systems,
\ben
G_{{\rm NW}}(\nu_{{\rm f}},\phi_{{\rm f}};\nu_{{\rm i}},\phi_{{\rm i}})\equiv\int\limits_{-\infty}^{\infty}d\alpha\; \langle\nu_{{\rm f}},\phi_{{\rm f}}|e^{\im\alpha\hat{\mathfrak{C}}}2|\hat{p}_{\phi}||\nu_{{\rm i}},\phi_{{\rm i}}\rangle\,,
\een
one splits $\alpha$ into $N$ parts and characterises each history $(\nu_{{\rm i}},\bar\nu_{1},\ldots,\nu_{{\rm f}})$ by the number of volume transitions. Then the limit $N\rightarrow\infty$ can be taken to yield the sum-over-histories form \cite{ach} \footnote{Notation: $\Theta_{\nu_i\nu_j}=\langle \nu_i|\Theta|\nu_j\rangle$ are matrix elements of $\Theta$; in the second line of (\ref{spinex}) the distinct values appearing in $(\nu_\im,\nu_1,\ldots,\nu_{M-1},\nu_{{\rm f}})$ are labelled by $w_1,\ldots,w_p$, with multiplicities $n_1,\ldots,n_p$ so that $\sum n_k=M+1$.}
\bea
G_{{\rm NW}}(\nu_{{\rm f}},\phi_{{\rm f}};\nu_{{\rm i}},\phi_{{\rm i}})&=&\sum_{M=0}^{\infty} \sum_{{\nu_{M-1},\ldots,\nu_1}\atop{\nu_m\neq \nu_{m+1}}}\Theta_{\nu_{{\rm f}}\nu_{M-1}}\ldots\Theta_{\nu_2 \nu_1}\Theta_{\nu_1 \nu_{{\rm i}}}\nn
\\&&\times\prod_{k=1}^p\frac{1}{(n_k-1)!}\left(\frac{\partial}{\partial \Theta_{w_k w_k}}\right)^{n_k-1}\sum_{m=1}^p\frac{2\cos(\sqrt{\Theta_{w_m w_m}}(\phi_{{\rm f}}-\phi_\im))}{\prod_{{j=1}\atop{j\neq m}}^p(\Theta_{w_m w_m}-\Theta_{w_j w_j})}\,.
\label{spinex}
\eea

Alternatively, one can pass to a deparametrised formalism, considering transition amplitudes $G_{{\rm NW}}^+(\nu_{{\rm f}},\phi_{{\rm f}};\nu_{{\rm i}},\phi_{{\rm i}})=\langle \nu_{{\rm f}}|e^{{\rm i}\sqrt{\Theta}(\phi_{{\rm f}}-\phi_{{\rm i}})}|\nu_{{\rm i}}\rangle$ for the ``square root" of (\ref{const}), $-{\rm i}\partial_{\phi}\Psi(\nu,\phi)=\sqrt{\Theta}\Psi(\nu,\phi)$,
\bea G^+_{{\rm NW}}(\nu_{{\rm f}},\phi_{{\rm f}};\nu_{{\rm i}},\phi_{{\rm i}})&=&\sum_{M=0}^{\infty} \sum_{{\nu_{M-1},\ldots,\nu_1}\atop{\nu_m\neq \nu_{m+1}}}(\sqrt{\Theta})_{\nu_{{\rm f}}\nu_{M-1}}\ldots(\sqrt{\Theta})_{\nu_2 \nu_1}(\sqrt{\Theta})_{\nu_1 \nu_{{\rm i}}}\times
\label{spinex2}
\\&&\prod_{k=1}^p\frac{1}{(n_k-1)!}\left(\frac{\partial}{\partial (\sqrt{\Theta})_{w_k w_k}}\right)^{n_k-1}\sum_{m=1}^p\frac{e^{{\rm i}(\sqrt{\Theta})_{w_m w_m}(\phi_{{\rm f}}-\phi_\im)}}{\prod_{{j=1}\atop{j\neq m}}^p((\sqrt{\Theta})_{w_m w_m}-(\sqrt{\Theta})_{w_j w_j})}\,.\nn
\eea
Our notation for the different definitions should become clear in the next section. (\ref{spinex}) and (\ref{spinex2}) define two different expansions of the same function (after restricting to positive frequency in (\ref{spinex}))\footnote{We show directly in \cite{paper} that they are the same function, the positive-frequency Newton-Wigner function $G_{{\rm NW}}^+$.}. Are there other possible definitions for the inner product? Do (\ref{spinex}) and (\ref{spinex2}) have good composition properties? In quantum gravity one might want to assume a relation like $G[g,g']=\int\mathcal{D}g''\,G[g,g'']G[g'',g']
$
which can be checked in this model for all possible two-point functions. We will see that all of them satisfy the same composition laws as for the relativistic particle. For details of all calculations see \cite{paper}.

\section{Two-point functions}
\subsection{Particle analogy}
First we detail the two-point functions for the relativistic particle following \cite{hallort}. Writing the relativistic particle in parametrised form $S=\int d\tau\;\left(p_a\dot{x}^a + N(p_x^2-p_t^2+m^2)\right)$ and fixing $\dot{N}=0$, one can define two-point functions through a sum-over-histories, such as the {\it Hadamard function}
\ben
G_{{\rm H}}(x'',t'';x',t')=\int\limits_{-\infty}^{\infty}dT\;g(x'',t'';T|x',t';0)\,,
\label{groupav}
\een
where $g(x'',t'';T|x',t';0)$ is a non-relativistic transition amplitude for $H=p_x^2-p_t^2+m^2$ in proper time $T$. Restricting the range of integration of proper time one obtains the {\it Feynman propagator}
\ben
\im G_{{\rm F}}(x'',t'';x',t')=\int\limits_{0}^{\infty}dT\;g(x'',t'';T|x',t';0)\,,
\label{feynm}
\een
which is a proper Green's function: $(\Box_{x'}-m^2)G_{{\rm F}}(x'',t'';x',t')=-\delta(x''-x';t''-t')$.
It satisfies a relativistic composition law involving a normal derivative,
\ben
G_{{\rm F}}(x'',t'';x',t')=-\int_{\Sigma}ds\;G_{{\rm F}}(x'',t'';x(s),t(s))\left(\stackrel{\rightarrow}{\partial_n}-\stackrel{\leftarrow}{\partial_n}\right)G_{{\rm F}}(x(s),t(s);x',t')\,,
\een
which is absent for the Hadamard function: $G_{{\rm H}}\circ G_{{\rm H}}=G_{{\rm c}}$, the causal two-point function \cite{hallort}.

Other two-point functions which require a splitting into positive and negative frequency are the Wightman functions $G^{\pm}$, so that $G_{{\rm H}}=G^++G^-$ and $\im G_{{\rm F}}=G^+\cdot\theta(t_{{\rm f}}-t_\im)+ G^-\cdot\theta(t_\im-t_{{\rm f}}),$ and the non-Lorentz invariant {\it Newton-Wigner function}
\ben
G^+_{{\rm NW}}(x'',t'';x',t')=\langle x''|e^{-\im \sqrt{m^2+k^2} (t''-t')}| x'\rangle=\int \frac{dk}{2\pi}e^{\im k(x''-x')}e^{-\im \sqrt{m^2+k^2} (t''-t')}
\een
which satisfies the non-relativistic composition law $G^+_{{\rm NW}}(x'',t'';x',t')=\int dx\; G^+_{{\rm NW}}(x'',t'';x,t)\times$ $G^+_{{\rm NW}}(x,t;x',t').$ The similarity of these expressions to quantum cosmology is obvious. The role of Lorentz invariance in minisuperspace is however rather unclear. 

\subsection{(Loop) quantum cosmology: Definitions and composition properties}

We can now derive vertex expansions for all two-point functions for quantum cosmology; the Hadamard function and Feynman propagator can be defined without deparametrisation. The analogue of (\ref{groupav}), $G_{{\rm H}}(\nu_{{\rm f}},\phi_{{\rm f}};\nu_{{\rm i}},\phi_{{\rm i}})\equiv\int\limits_{-\infty}^{\infty}d\alpha\; \langle\nu_{{\rm f}},\phi_{{\rm f}}|e^{\im\alpha\hat{\mathfrak{C}}}|\nu_{{\rm i}},\phi_{{\rm i}}\rangle$, gives the Hadamard function
\bea
G_{{\rm H}}(\nu_{{\rm f}},\phi_{{\rm f}};\nu_{{\rm i}},\phi_{{\rm i}})&=&\sum_{M=0}^{\infty} \sum_{{\nu_{M-1},\ldots,\nu_1}\atop{\nu_m\neq \nu_{m+1}}}\Theta_{\nu_{{\rm f}}\nu_{M-1}}\ldots\Theta_{\nu_2 \nu_1}\Theta_{\nu_1 \nu_{{\rm i}}}\times
\\&&\prod_{k=1}^p\frac{1}{(n_k-1)!}\left(\frac{\partial}{\partial \Theta_{w_k w_k}}\right)^{n_k-1}\sum_{m=1}^p\frac{1}{\sqrt{\Theta_{w_m w_m}}}\frac{\cos(\sqrt{\Theta_{w_m w_m}}(\phi_{{\rm f}}-\phi_\im))}{\prod_{{j=1}\atop{j\neq m}}^p(\Theta_{w_m w_m}-\Theta_{w_j w_j})}\,.\nn
\eea
Analogously, one derives the Feynman propagator by restricting the range of integration to positive $\alpha$, as in (\ref{feynm}), and following the ``$\im\epsilon$" contour in the complex $p_\phi$ plane to get
\bea
\im G_{{\rm F}}(\nu_{{\rm f}},\phi_{{\rm f}};\nu_{{\rm i}},\phi_{{\rm i}})&=&\sum_{M=0}^{\infty} \sum_{{\nu_{M-1},\ldots,\nu_1}\atop{\nu_m\neq \nu_{m+1}}}\Theta_{\nu_{{\rm f}}\nu_{M-1}}\ldots\Theta_{\nu_2 \nu_1}\Theta_{\nu_1 \nu_{{\rm i}}}\prod_{k=1}^p\frac{1}{(n_k-1)!}\times
\\&&\left(\frac{\partial}{\partial \Theta_{w_k w_k}}\right)^{n_k-1}\sum_{m=1}^p\frac{e^{-\im\sqrt{\Theta_{w_m w_m}}\Delta\phi}\theta(\Delta\phi)+e^{\im\sqrt{\Theta_{w_m w_m}}\Delta\phi}\theta(-\Delta\phi)}{2 \sqrt{\Theta_{w_m w_m}} \prod_{{j=1}\atop{j\neq m}}^p(\Theta_{w_m w_m}-\Theta_{w_j w_j})}\nn
\eea
with $\Delta\phi\equiv\phi_{{\rm f}}-\phi_{{\rm i}}$. The $\phi$ dependence of these expressions is analogous to the two-point functions for the relativistic particle; they also satisfy the same composition laws.
\\First one can show that the Newton-Wigner function restricted to the positive frequency sector, 
\ben
G_{{\rm NW}}^+(\nu_{{\rm f}},\phi_{{\rm f}};\nu_\im,\phi_\im)\equiv\int\limits_{-\infty}^{\infty}d\alpha\int\limits_0^{\infty} \frac{dp_\phi}{2\pi}\,\langle\nu_{{\rm f}}|e^{-\im\alpha\Theta}|\nu_\im\rangle (2|p_\phi|) e^{\im\alpha p_\phi^2}e^{\im p_\phi \Delta\phi}=\langle\nu_{{\rm f}}|e^{\im\sqrt{\Theta}\,\Delta\phi}|\nu_\im\rangle\,,
\een
satisfies the usual non-relativistic composition law
\ben
G_{{\rm NW}}^+(\nu_{{\rm f}},\phi_{{\rm f}};\nu_\im,\phi_\im) = \sum_\nu G_{{\rm NW}}^+(\nu_{{\rm f}},\phi_{{\rm f}};\nu,\phi) G_{{\rm NW}}^+(\nu,\phi;\nu_\im,\phi_\im)\,.
\een
There is no composition law if one includes both positive- and negative-frequency sectors. It is essential to be able to separate these sectors. 

For the Hadamard function, one has, just as for the relativistic particle,
\ben
G_{{\rm c}}(\nu_{{\rm f}},\phi_{{\rm f}};\nu_\im,\phi_\im) = \sum_\nu G_{{\rm H}}(\nu_{{\rm f}},\phi_{{\rm f}};\nu,\phi) \left(\stackrel{\rightarrow}{\partial_\phi} - \stackrel{\leftarrow}{\partial_\phi}\right)G_{{\rm H}}(\nu,\phi;\nu_\im,\phi_\im)
\een
because of different composition law for positive and negative frequency Wightman functions.

Finally, the Feynman propagator satisfies, again in agreement with the relativistic particle,
\ben
\im G_{{\rm F}}(\nu_{{\rm f}},\phi_{{\rm f}};\nu_\im,\phi_\im) = \im\sum_\nu \im G_{{\rm F}}(\nu_{{\rm f}},\phi_{{\rm f}};\nu,\phi) \left(\stackrel{\rightarrow}{\partial_\phi} - \stackrel{\leftarrow}{\partial_\phi}\right)\im G_{{\rm F}}(\nu,\phi;\nu_\im,\phi_\im)\,.
\een
Recall that it is not a projector on solutions to the constraint $\mathfrak{C}$, but a proper Green's function.

\section{Summary and Outlook}
Exploiting the analogy with the relativistic particle, we have given explicit expressions for all two-point functions (Hadamard, Feynman, Newton-Wigner, etc), for constrained dynamics of the form $\mathfrak{C}=p_\phi^2-\Theta$. They satisfy the same composition properties as their analogues. For the definition of some of the two-point functions the existence of an explicit splitting into positive and negative frequency is essential; such a splitting is not generally available for spin foams or more general cosmological models. Without the frequency splitting there is no two-point function defining a physical inner product (i.e. a projector on solutions to the constraint) and satisfying a ``nice" composition law. Possible issues with the composition properties of the usual definition of the inner product through spin foams have been discussed as the ``cosine problem", which is closely related to the different composition laws for positive and negative frequency Wightman functions that we have seen above.

In our model, the splitting of solutions into two sectors is of course possible and for a single particle (=Universe) everything is consistent when one restricts to one sector. The different choices of two-point functions become relevant in a field theory (third quantisation) picture, where one might follow an argument by Kucha\v{r} \cite{kuchar} for geometrodynamics: If one needs a splitting analogous to positive and negative frequency to have a consistent one-Universe quantum mechanics, one would require a conserved quantity analogous to energy, corresponding to a Killing vector, on (metric) superspace, which however does not exist. If one takes this argument (which is only based on an analogy) seriously, one has to consider the equivalent of QFT on curved spacetime, formulated without the fundamental concept of a single particle. 

For the cosmological model at hand, this reasoning leads naturally to consider a GFT-like model of a quantum field theory on the 2-dimensional space spanned by $\nu$ and $\phi$, where one could add interactions to implement topology change. One possible physical interpretation of that would be the creation of inhomogeneities; other questions addressed in this framework include the role of the ``GFT coupling constant" $\lambda$. Work on this is currently in progress.


\section*{References}
\begin{thebibliography}{42}
\bibitem{talkpdf} \verb"http://loops11.iem.csic.es/loops11/index.php?option=com_content&view=article&id=75&Itemid=73"
\bibitem{ach} Ashtekar A, Campiglia M and Henderson A 2009 Loop Quantum Cosmology and Spin Foams {\it Physics Letters} {\bf B681} 347--352; Ashtekar A, Campiglia M and Henderson A 2010 Casting Loop Quantum Cosmology in the Spin Foam Paradigm {\it Classical and Quantum Gravity} {\bf 27} 135020 
\bibitem{paper} Calcagni G, Gielen S and Oriti D 2011 Two-point functions in (loop) quantum cosmology {\it Classical and Quantum Gravity} {\bf 28} 125014
\bibitem{hallort} Halliwell JJ and Ortiz ME 1993 Sum-over-histories origin of the composition laws of relativistic quantum mechanics and quantum cosmology {\it Physical Review} {\bf D48} 748--768
\bibitem{kuchar} Kucha\v{r} K 1981 General relativity: Dynamics without symmetry {\it Journal of Mathematical Physics} {\bf 22} 2640--2654

\end{thebibliography}
\end{document}